\documentclass[11pt]{article}

\usepackage[truedimen,margin=25truemm]{geometry}

\usepackage{kasai_def}

\usepackage{graphicx}

\usepackage{listings}
\usepackage{color}

\definecolor{codegreen}{rgb}{0,0.6,0}
\definecolor{codegray}{rgb}{0.5,0.5,0.5}
\definecolor{codepurple}{rgb}{0.58,0,0.82}
\definecolor{backcolour}{rgb}{0.95,0.95,0.92}

\lstdefinestyle{mystyle}{
  backgroundcolor=\color{backcolour},   
  commentstyle=\color{codegreen},
  keywordstyle=\color{magenta},
  numberstyle=\tiny\color{codegray},
  stringstyle=\color{codepurple},
  basicstyle=\footnotesize\ttfamily,
  breakatwhitespace=false,         
  breaklines=true,                 
  captionpos=b,                    
  keepspaces=true,                 
  numbers=left,                    
  numbersep=5pt,                  
  showspaces=false,                
  showstringspaces=false,
  showtabs=false,                  
  tabsize=2,
  frame=single
}

\usepackage{dirtree}

\newcommand{\changeHK}[1]{\textcolor{black}{#1}}
\newcommand{\changeHKK}[1]{\textcolor{black}{#1}}

\title{SGDLibrary:\\ A MATLAB library for stochastic gradient descent algorithms}
\date{\today\\First version: October 27, 2017 }

\author{Hiroyuki Kasai\thanks{Graduate School of Informatics and Engineering, The University of Electro-Communications, Tokyo, Japan ({\tt kasai@is.uec.ac.jp}).} }

\begin{document}

\maketitle

\begin{abstract}
We consider the problem of finding the minimizer of a function $f: \mathbb{R}^d \rightarrow \mathbb{R}$ of the finite-sum form $\min f(w) = 1/n\sum_{i}^n f_i(w)$. This problem has been studied intensively in recent years in the field of machine learning (ML). One promising approach for large-scale data is to use a stochastic optimization algorithm to solve the problem. SGDLibrary is a readable, flexible and extensible pure-MATLAB library of a collection of stochastic optimization algorithms. The purpose of the library is to provide researchers and implementers a comprehensive evaluation environment for the use of these algorithms on various ML problems.
\end{abstract}

\begin{center}
\vspace*{0.3cm}
\textcolor{blue}{
{\scriptsize
Published in Journal of Machine Learning Research (JMLR) entitled\\
``SGDLibrary: A MATLAB library for stochastic gradient optimization algorithms"  \cite{Kasai_JMLR_2018_s}
}
}
\end{center}


%

\section{Introduction}
This work aims to facilitate research on stochastic optimization for large-scale data. We particularly address a regularized finite-sum minimization problem defined as 
\begin{eqnarray}
	\label{Eq:ProblemDefinition}
	\min_{w \in \mathbb{R}^d} f(w) := \frac{1}{n} \sum_{i=1}^n f_i(w) 
	= \frac{1}{n} \sum_{i=1}^n L(w, x_i, y_i)  + \lambda R(w),
\end{eqnarray}
where $w \in \mathbb{R}^d$ represents the model parameter and $n$ denotes the number of samples \changeHK{$(x_i, y_i)$}. $L(w, x_i, y_i)$ is the loss function and $R(w)$ is the regularizer with the regularization parameter $\lambda\geq 0$. Widely diverse machine learning (ML) models fall into this problem. Considering $L(w, x_i, y_i) = (w^Tx_i- y_i)^2$, $x_i \in \mathbb{R}^d$, $y_i \in \mathbb{R}$ and $R(w)=\| w \|_2^2$, this results in \changeHKK{an} $\ell_2$-norm regularized linear regression problem (a.k.a. ridge regression) for $n$ training samples $(x_1, y_1), \cdots, (\changeHK{x_n}, y_n)$. 
In \changeHKK{the} case of binary \changeHKK{classification with} the desired class label $y_i \in \{+1,-1\}$ and $R(w)=\| w\|_1$, \changeHKK{an} $\ell_1$-norm regularized logistic regression (LR) problem is obtained as $f_i(w)=\log (1+\exp(-y_i w^Tx_i )) +  \lambda \| w \|_1$, which encourages the sparsity of the solution of $w$. Other problems \changeHKK{covered} are matrix completion, support vector machines (SVM), and sparse principal components analysis, to name but a few. 

{\it Full gradient \changeHK{descent}} (a.k.a. steepest descent) with a step-size $\eta$ is the most straightforward approach for (\ref{Eq:ProblemDefinition}),
\changeHK{which updates as $w_{k+1}  \leftarrow  w_{k} - \changeHKK{\eta} \nabla f(w_{k})$ at \changeHKK{the} $k$-th iteration}.
However, this is expensive when $n$ is extremely large. In fact, one needs a sum of $n$ calculations of the inner product $w^Tx_i$ 
\changeHK{in the regression problems above}, leading to $\mathcal{O}(nd)$ cost overall per iteration. 
For this issue, a popular and effective alternative is 
{\it stochastic gradient descent} (SGD), which updates as $w_{k+1} \leftarrow  w_{k} - \changeHKK{\eta}\nabla f_i(w_{k})$ for \changeHKK{the} $\changeHKK{i}$-th sample uniformly at random \changeHK{\cite{Robbins_MathStat_1951,Bottou_CUP_1998}}. SGD assumes an {\it unbiased estimator} of the full gradient as $\mathbb{E}_i[\nabla f_i(w^k)] = \nabla f (w^k)$. As the update rule represents, the calculation cost is independent of $n$, resulting in $\mathcal{O}(d)$ \changeHK{per iteration}. \changeHK{Furthermore,} {\it mini-batch} SGD \changeHK{\cite{Bottou_CUP_1998}} calculates $1/|\mathcal{S}_k| \sum_{i \in \mathcal{S}_k} \nabla f_i(w_{k})$, where $\mathcal{S}_k$ is the set of samples of size $|\mathcal{S}_k|$. SGD needs a {\it diminishing} step-size algorithm to guarantee \changeHKK{convergence}, which causes a severe slow convergence rate \changeHK{\cite{Bottou_CUP_1998}}. 
To accelerate this rate, we have two active research directions in ML; 
{\it Variance reduction} (VR) techniques \cite{Johnson_NIPS_2013_s_new,Roux_NIPS_2012_s_new,Shalev_JMLR_2013_s,Defazio_NIPS_2014_s,Nguyen_ICML_2017} \changeHKK{that} explicitly or implicitly exploit a full gradient estimation to reduce the variance of \changeHKK{the} noisy stochastic gradient, leading to superior convergence properties. 
%
Another promising direction is to modify deterministic {\it second-order} algorithms into stochastic settings, and \changeHKK{solve} the potential problem of {\it first-order} algorithms for {\it ill-conditioned} problems. A direct extension of {\it quasi-Newton} (QN) is known as online BFGS \cite{Schraudolph_AISTATS_2007_s}. Its variants include \changeHKK{a} regularized version (RES) \cite{Mokhtari_IEEETranSigPro_2014}, \changeHKK{a} limited memory version (oLBFGS) \cite{Schraudolph_AISTATS_2007_s,Mokhtari_JMLR_2015_s}, \changeHKK{a} stochastic QN (SQN) \cite{Byrd_SIOPT_2016}, \changeHKK{an} incremental QN \cite{Mokhtari_ICASSP_2017}, and \changeHKK{a} non-convex version. Lastly, hybrid algorithms of the SQN  with VR are proposed \cite{Moritz_AISTATS_2016_s_new,Kolte_OPT_2015}. Others include \cite{Duchi_JMLR_2011_abb,Bordes_JMLR_2009_abb}.

The performance of stochastic optimization algorithms is strongly influenced not only by the distribution of data but also by the step-size algorithm \changeHK{\cite{Bottou_CUP_1998}}.
Therefore, we often encounter results that are completely different from those in papers in every experiment. Consequently, an evaluation framework to test and compare the algorithms at hand is crucially important for fair and comprehensive experiments. \changeHK{One existing tool is {\it Lightning} \cite{lightning_2016}, which is a Python library for large-scale ML problems. However, its supported algorithms are limited, and the solvers and the problems such as classifiers are mutually connected. Moreover, the implementations utilize Cython, which is a C-\changeHKK{extension} for Python, for efficiency. Subsequently, they decrease users' readability of \changeHKK{code}, and also make users' evaluations and extensions more complicated.} SGDLibrary is a readable, flexible and extensible pure-\changeHK{MATLAB} library of a collection of stochastic optimization algorithms. \changeHK{The library is also operable on GNU Octave.} The purpose of the library is to provide researchers and implementers a collection of \changeHK{state-of-the-art} stochastic optimization algorithms that solve a variety of large-scale optimization problems such as linear/non-linear regression problems and classification problems. 
\changeHK{This also allows researchers and implementers to easily extend or add solvers and problems for further \changeHKK{evaluation}.} 
To the best of my knowledge, no report in the literature \changeHK{and no library} describe a comprehensive experimental environment specialized for stochastic optimization algorithms. The code is available \changeHKK{at} \url{https://github.com/hiroyuki-kasai/SGDLibrary}.

\section{Software architecture}
The software architecture of SGDLibrary follows a typical {\it module-based} architecture, which separates {\it problem descriptor} and {\it optimization solver}. To use the library, the user selects one problem descriptor of interest and no less than one optimization solvers to be compared. 

\noindent {\bf Problem descriptor:}
The problem descriptor, denoted as \verb+problem+, specifies the problem of interest with respect to $w$, noted as \verb+w+ in the library. \changeHK{This is implemented by MATLAB} \verb+classdef+. The user does nothing other than calling a problem definition function, for instance, \verb+logistic_regression()+ for \changeHKK{the} $\ell_2$-norm regularized LR problem. Each problem definition includes the functions necessary for solvers;
(i) (full) cost function $f(w)$,
(ii) 
mini-batch stochastic derivative  \verb+v+=$1/|\mathcal{S}| \nabla f_{i \in \mathcal{S}} (w)$ for the set of samples $\mathcal{S}$.
(iii)
stochastic Hessian \changeHK{\cite{Bordes_JMLR_2009_abb}}, 
and
(iv) 
stochastic Hessian-vector product for a vector \verb+v+.
The \changeHK{built-in} problems include\changeHK{, for example,} $\ell_2$-norm regularized multidimensional linear regression, $\ell_2$-norm regularized linear SVM, $\ell_2$-norm regularized LR, $\ell_2$-norm regularized softmax classification (multinomial LR), $\ell_1$-norm multidimensional linear regression, and $\ell_1$-norm LR. The problem descriptor provides additional specific functions. For example, the LR problem includes the prediction and the classification accuracy calculation functions. 

\noindent {\bf Optimization solver:}
The optimization solver implements the main routine of the stochastic optimization algorithm. Once a solver function is called with one selected problem descriptor \verb+problem+ as the first argument, it solves the optimization problem by calling some corresponding functions via \verb+problem+ such as the cost function and the stochastic gradient calculation function. 
\changeHK{\changeHKK{Examples of} the} supported optimization solvers in the library are listed \changeHKK{in} categorized groups as;
\noindent (i) {\bf SGD methods:} Vanila SGD \cite{Robbins_MathStat_1951}, SGD with classical momentum, SGD with classical momentum with Nesterov's accelerated gradient \cite{Sutskever_ICML_2013}, AdaGrad \cite{Duchi_JMLR_2011_abb}, RMSProp, AdaDelta, Adam, and AdaMax,
 (ii) {\bf Variance reduction (VR) methods:} SVRG \cite{Johnson_NIPS_2013_s_new}, SAG \cite{Roux_NIPS_2012_s_new}, SAGA \cite{Defazio_NIPS_2014_s}, and SARAH \cite{Nguyen_ICML_2017},
(iii) {\bf Second-order methods}: SQN \cite{Bordes_JMLR_2009_abb}, oBFGS-Inf \cite{Schraudolph_AISTATS_2007_s, Mokhtari_JMLR_2015_s},
oBFGS-Lim (oLBFGS) \cite{Schraudolph_AISTATS_2007_s, Mokhtari_JMLR_2015_s},
Reg-oBFGS-Inf (RES) \cite{Mokhtari_IEEETranSigPro_2014}, and Damp-oBFGS-Inf,
(iv)
{\bf Second-order method\changeHKK{s} with VR}: SVRG-LBFGS \cite{Kolte_OPT_2015}, SS-SVRG \cite{Kolte_OPT_2015}, and SVRG-SQN \cite{Moritz_AISTATS_2016_s_new}, and
(v)
\noindent {\bf Else}: BB-SGD and SVRG-BB.
The solver \changeHK{function} also receives optional parameters as the second argument, which forms a {\it struct}, designated as \verb+options+ in the library. It contains elements such as the maximum number of epochs, the batch size, and the step-size algorithm with an initial step-size. Finally, the solver \changeHK{function} returns to the caller the final solution \verb+w+ and rich \changeHKK{statistical} information, \changeHKK{such as a record} of the cost function value\changeHKK{s}, \changeHKK{the} optimality gap, \changeHK{\changeHKK{the} processing time,} and the number of gradient calculations. 

\noindent{\bf Others:}
SGDLibrary accommodates a {\it user-defined} step-size algorithm. This accommodation is achieved by setting as \verb+options.stepsizefun=@my_stepsize_alg+, which is delivered to solvers. Additionally, when the regularizer $R(w)$ in the minimization problem ({\ref{Eq:ProblemDefinition}) is a non-smooth regularizer such as \changeHKK{the} $\ell_1$-norm regularizer $\| w\|_1$, the solver calls the {\it proximal operator} as \verb+problem.prox(w,stepsize)+, which is the wrapper function defined in each problem. \changeHKK{The} $\ell_1$-norm regularized LR problem, for example, calls \changeHKK{the} {\it soft-threshold} function as \verb+w = prox(w,stepsize)=soft_thresh(w,stepsize*lambda)+, where \verb+stepsize+ is the step-size $\eta$ and \verb+lambda+ is the regularization parameter $\lambda > 0$ in (\ref{Eq:ProblemDefinition}).

\section{Tour of the SGDLibrary}

We embark on a tour of SGDLibrary exemplifying the $\ell_2$-norm regularized LR problem. The LR model generates $n$ pairs of $(x_i, y_i)$ for an unknown model parameter $w$, where $x_i$ is an input $d$-dimensional vector and $y_i \in \{-1, 1\}$ is the binary class label,  as 
$
P(y_i | x_i, w)  =  1/(1 + \exp(-y_i w^T x_i)).
$
The problem seeks $w$ that fits the regularized LR model to the generated data $(x_i, y_i)$. This problem is cast as a minimization problem as
$
\min f(w)  := 1/n \sum_{i=1}^n \log [1 + \exp(-y_i w^T x_i)] + \lambda/2 \| w \|^2. 
$
The code for this problem is in Listing \ref{demo}.
\begin{lstlisting}[language=Matlab, label=demo, caption=Demonstration code for logistic regression problem.]
% generate synthetic 300 samples of dimension 3 for logistic regression
d = logistic_regression_data_generator(300,3);
% define logistic regression problem
problem = logistic_regression(d.x_train,d.y_train,d.x_test,d.y_test); 

options.w_init = d.w_init;                      % set initial value  
options.step_init = 0.01;                       % set initial stepsize
options.verbose = 1;                            % set verbose mode
[w_sgd, info_sgd] = sgd(problem, options);      % perform SGD solver
[w_svrg, info_svrg] = svrg(problem, options);   % perform SVRG solver
[w_svrg, info_svrg] = sqn(problem, options);   % perform SQN solver
% display cost vs. number of gradient evaluations
display_graph('grad_calc_count','cost',{'SGD','SVRG'},...
                {w_sgd,w_svrg},{info_sgd,info_svrg});
\end{lstlisting}

First, we generate train/test datasets \verb+d+ using \verb+logistic_regression_data_generator()+, where the input feature vector is with $n=300$ and $d=3$. $y_i \in \{-1, 1\}$ is its class label. The LR problem is defined by calling \verb+logistic_regression()+, which internally contains the functions for cost value, the gradient and the Hessian. This is stored in \verb+problem+. Then, we execute solvers, i.e., SGD and SVRG, by calling solver functions, i.e., \verb+sgd()+ and \verb+svrg()+  with \verb+problem+ and \verb+options+ after setting some options into the \verb+options+ struct. They return the final solutions of \verb+{w_sgd,w_svrg}+ and the \changeHKK{statistical} information \verb+{info_sgd,info_svrg}+. Finally, \verb+display_graph()+ visualizes the behavior of the cost function values in terms of the number of gradient evaluations. It is noteworthy that each algorithm requires a different number of evaluations of samples in each epoch. Therefore, it is common to use this value to evaluate the algorithms instead of the number of iterations. \changeHKK{Illustrative} results \changeHK{additionally including SQN and SVRG-LBFGS} are presented in Figure \ref{fig:example_of_results}, which are generated by \verb+display_graph()+, and  \verb+display_classification_result()+ specialized for classification problems.  
Thus, SGDLibrary provides rich visualization tools as well.
\begin{figure}[htbp]
\vspace*{-0.2cm}
\begin{center}
\begin{minipage}{0.32\hsize}
\begin{center}
\includegraphics[width=1.05\hsize]{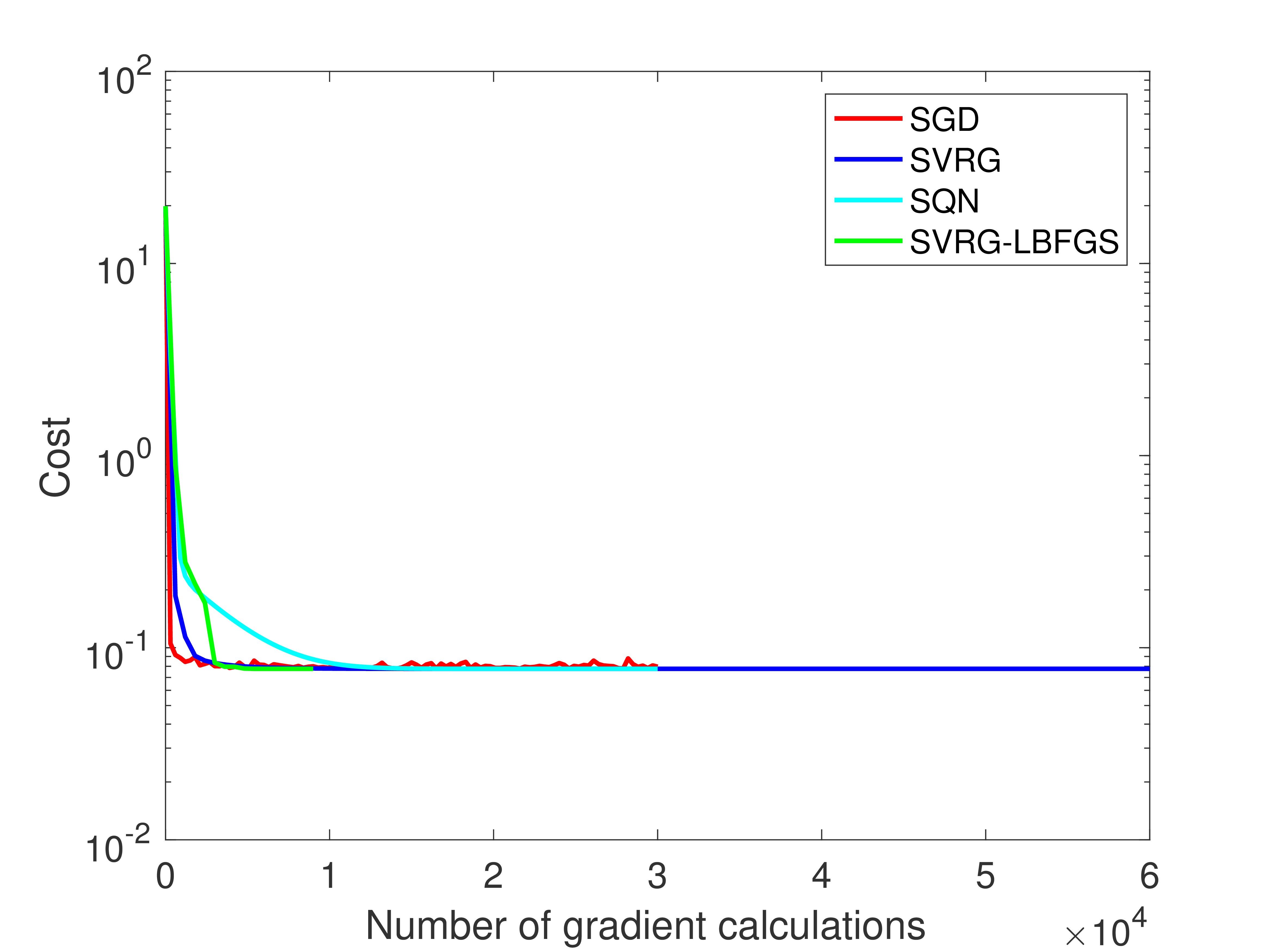}
{\scriptsize (a) Cost function value}
\end{center}
\end{minipage}
\begin{minipage}{0.32\hsize}
\begin{center}
\includegraphics[width=1.05\hsize]{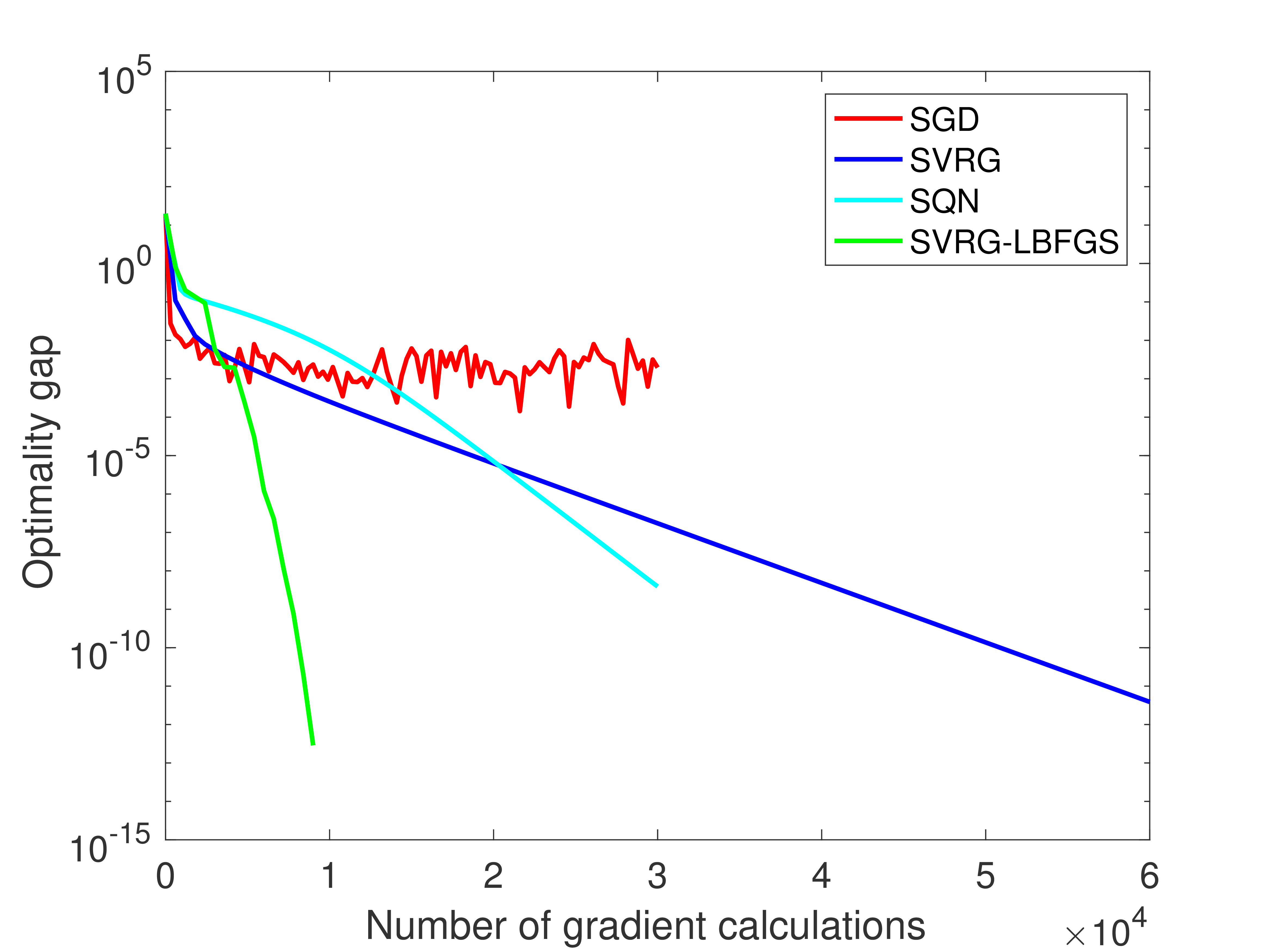}
{\scriptsize (b) Optimality gap}
\end{center}
\end{minipage}
\begin{minipage}{0.32\hsize}
\begin{center}
\includegraphics[width=1.05\hsize]{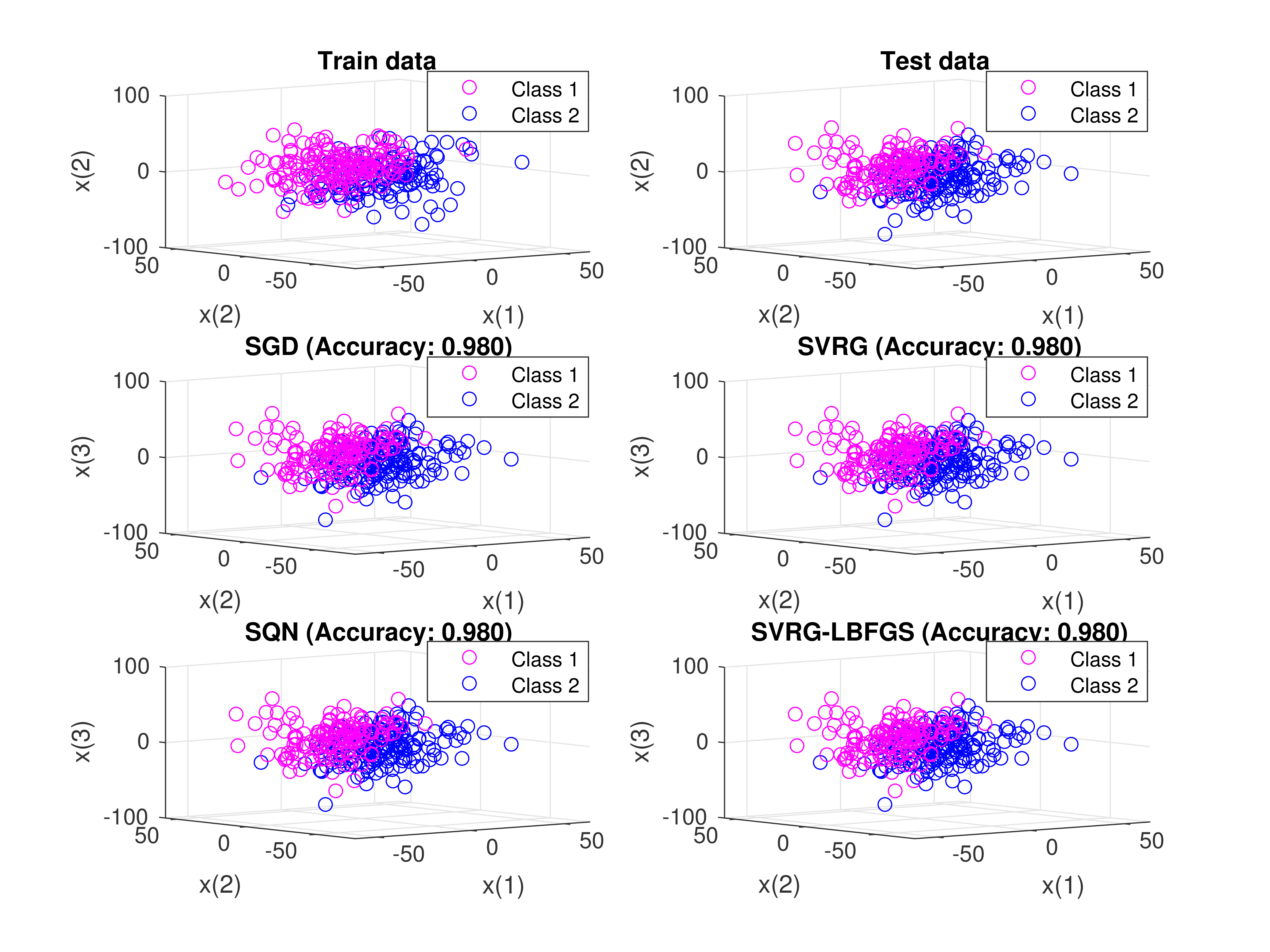}
{\scriptsize (c) Classification result}
\end{center}
\end{minipage}
\vspace*{-0.0cm}
\caption{Results of $\ell_2$-norm regularized logistic regression problem.}

\label{fig:example_of_results}
\end{center}
\vspace*{-1.5cm}
\end{figure}

\clearpage
\appendix
\section{Supported stochastic optimization solvers}

Table \ref{tbl:Algorithms} summarizes the supported stochastic optimization algorithms and configurations.

\begin{table}[htbp]
\caption{Supported stochastic optimization algorithms and configurations}
\label{tbl:Algorithms}
\begin{center}
\begin{tabular}{l|l|l|l|l}
\hline
algorithm name& solver & \verb+sub_mode+ & other \verb+options+ & Reference\\
\hline
\hline
SGD&\verb+sgd.m+& &&\cite{Robbins_MathStat_1951}\\
\hline
SGD-CM&\verb+sgd_cm.m+&\verb+CM+   &&\\
\hline
SGD-CM-NAG&\verb+sgd_cm.m+&\verb+CM-NAG+&& \cite{Sutskever_ICML_2013}\\
\hline
AdaGrad&\verb+adagrad.m+&\verb+AdaGrad+&&\cite{Duchi_JMLR_2011_abb}\\
\hline
RMSProp & \verb+adagrad.m+ &\verb+RMSProp+&&\cite{Tieleman_2012}\\
\hline
AdaDelta&\verb+adagrad.m+&\verb+AdaDelta+&&\cite{Zeiler_arXiv_2012}\\
\hline
Adam&\verb+adam.m+&\verb+Adam+&&\cite{Kingma_ICLR_2015}\\
\hline
AdaMax&\verb+adam.m+&\verb+AdaMax+&&\cite{Kingma_ICLR_2015}\\
\hline
\hline
SVRG&\verb+svrg.m+&&&\cite{Johnson_NIPS_2013_s}\\
\hline
SAG&\verb+sag.m+&\verb+SAG+&&\cite{Roux_NIPS_2012_s}\\
\hline
SAGA&\verb+sag.m+&\verb+SAGA+&&\cite{Defazio_NIPS_2014_s}\\
\hline
SARAH&\verb+sarah.m+&&&\cite{Nguyen_ICML_2017}\\
\hline
SARAH-Plus&\verb+sarah.m+&\verb+Plus+&&\cite{Nguyen_ICML_2017}\\
\hline
\hline
SQN&\verb+slbfgs.m+&\verb+SQN+&&\cite{Byrd_SIOPT_2016}\\
\hline
oBFGS-Inf&\verb+obfgs.m+&\verb+Inf-mem+&&\cite{Schraudolph_AISTATS_2007_s}\\
\hline
oLBFGS-Lim&\verb+obfgs.m+&\verb+Lim-mem+&&\cite{Schraudolph_AISTATS_2007_s}\\
&&&&\cite{Mokhtari_JMLR_2015_s}\\
\hline
Reg-oBFGS-Inf&\verb+obfgs.m+&\verb+Inf-mem+&\verb+delta+ $\neq 0$&\cite{Mokhtari_IEEETranSigPro_2014}\\
\hline
Damp-oBFGS-Inf& \verb+obfgs.m+ &\verb+Inf-mem+&\verb+delta+ $\neq 0$ \&&  \cite{Wang_SIOPT_2017}\\
& && \verb+damped=true+\\
\hline
IQN&\verb+iqn.m+&&&\cite{Mokhtari_ICASSP_2017}\\
\hline
\hline
SVRG-SQN&\verb+slbfgs.m+&\verb+SVRG-SQN+&&\cite{Moritz_AISTATS_2016_s}\\
\hline
SVRG-LBFGS&\verb+slbfgs.m+&\verb+SVRG-LBFGS+&&\cite{Kolte_OPT_2015}\\
\hline
SS-SVRG&\verb+subsamp+&&&\cite{Kolte_OPT_2015}\\
&\verb+_svrg.m+&&&\\
\hline
\hline
BB-SGD&\verb+bb_sgd.m+&&& \cite{De_AISTATS_2017}\\
\hline
SVRG-BB &\verb+svrg_bb.m+&&& \cite{Tan_NIPS_2016}\\
\hline
\end{tabular}
\end{center}
\end{table}

\clearpage
\section{Directory and file structure}

The tree structure below represents the directory and file structure of SGDLibrary. 
\renewcommand*\DTstylecomment{\color{black}}
\renewcommand*\DTstyle{\ttfamily\textcolor{blue}}
\dirtree{%
.1 /.
.2 README.md\DTcomment{Readme file}.
.2 run$\_$me$\_$first.m\DTcomment{First script to run}.
.2 demo.m\DTcomment{Demonstration script to check library}.
.2 demo$\_$ext.m\DTcomment{Demonstration script to check library}.
.2 sgdlibrary$\_$version.m\DTcomment{Version and release date information}.
.2 LICENSE.txt\DTcomment{License file}.
.2 problem/\DTcomment{Problem definitions}.
.2 sgd$\_$solver/\DTcomment{Stochastic optimization sovlers}.
.2 sgd$\_$test/\DTcomment{Test scripts to use this library}.
.2 plotter/\DTcomment{Tools for plotting}.
.2 tool/\DTcomment{Auxiliary tools}.
.2 gd$\_$solver/\DTcomment{Gradient descent optimization solver files}.
.2 gd$\_$test/\DTcomment{Test scripts for gradient descent solvers}.
}

\clearpage
\section{How to use SGDLibrary}

\subsection{First to do}

Run \verb+run_me_first+ for path configurations. 
\begin{verbatim}
>> run_me_first; 
##########################################################
###                                                    ###
###                Welcome to SGDLibrary               ###
###        (version:1.0.16, released:01-April-2017)    ###
###                                                    ###
##########################################################
\end{verbatim}

Now, we are ready to use the library. Just execute \verb+demo+ for the simplest demonstration of this library. This is the case of \L{2}-norm regularized logistic regression problem.

\begin{verbatim}
>> demo; 
SGD: Epoch = 000, cost = 3.3303581454622559e+01, optgap = Inf
SGD: Epoch = 001, cost = 1.1475439725295747e-01, optgap = Inf
............
SGD: Epoch = 099, cost = 7.9162101289075484e-02, optgap = Inf
SGD: Epoch = 100, cost = 7.8493133230913101e-02, optgap = Inf
Max epoch reached: max_epochr = 100
SVRG: Epoch = 000, cost = 3.3303581454622559e+01, optgap = Inf
SVRG: Epoch = 001, cost = 1.531453394302157711148737e-01, optgap = Inf
SVRG: Epoch = 002, cost = 9.330987606770735354189128e-02, optgap = Inf
............
SVRG: Epoch = 099, cost = 7.715265139028108787311311e-02, optgap = Inf
SVRG: Epoch = 100, cost = 7.715265135253832062822710e-02, optgap = Inf
Max epoch reached: max_epochr = 100
\end{verbatim}

The cost function values every iteration of two algorithms are shown in the Matlab command window. Additionally, the convergence plots of the cost function values are shown as in Figure \ref{fig:FirstDemoOutput}.

\begin{figure}[htbp]
\begin{center}
\includegraphics[width=8cm]{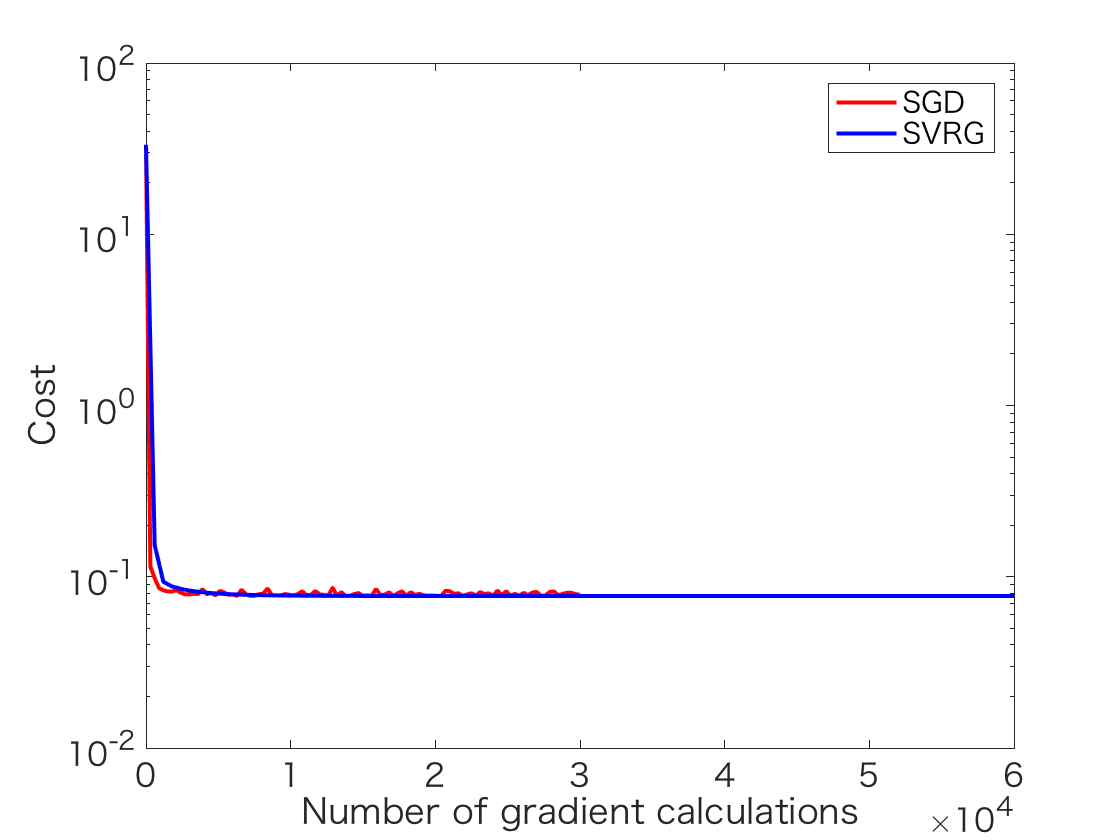}
\caption{Result of \L{2}-norm regularized logistic regression problem.} 
 \label{fig:FirstDemoOutput}
\end{center}
\end{figure}

\subsection{Simplest usage example: 4 steps!}

The code of demo.m is shown in Listing \ref{first_demo}.

\begin{lstlisting}[language=Matlab, label=first_demo, caption=Demonstration code for the logistic regression problem]
% generate synthetic 300 samples of dimension 3 for logistic regression
d = logistic_regression_data_generator(300,3);

% define logistic regression problem
problem = logistic_regression(d.x_train,d.y_train,d.x_test,d.y_test); 

options.w_init = d.w_init;                      % set initial value  
options.step_init = 0.01;                       % set initial stepsize
opitons.verbose = 2;                            % set verbose mode
[w_sgd, info_sgd] = sgd(problem, options);      % perform SGD solver
[w_svrg, info_svrg] = svrg(problem, options);   % perform SVRG solver

% display cost vs. number of gradient evaluations
display_graph('grad_calc_count','cost',{'SGD','SVRG'},...
                {w_sgd,w_svrg},{info_sgd,info_svrg});
\end{lstlisting}

Let us take a closer look at the code above bit by bit. The procedure has only 4 steps!\\

\noindent {\bf Step 1: Generate data}

First, we generate datasets including train set and test set using a data generator function \verb+logistic_regression_data_generator()+. The output includes train \& test set and an initial value of the solution \verb+w+.

\begin{lstlisting}[language=Matlab, label=demo_data_generate, caption=Code for data generation., firstnumber=1]
% generate synthetic 300 samples of dimension 3 for logistic regression
d = logistic_regression_data_generator(300,3);
\end{lstlisting}

\noindent {\bf Step 2: Define problem}

The problem to be solved should be defined properly from the supported problems. \verb+logistic_regression()+ provides the comprehensive functions for a logistic regression problem. This returns the cost value by \verb+cost(w)+, the gradient by \verb+grad(w)+ and the hessian by \verb+hess(w)+ when given w. These are essential for any gradient descent algorithms.

\begin{lstlisting}[language=Matlab, label=demo_problem_def, caption=Code for problem definition., firstnumber=4]
% define logistic regression problem
problem = logistic_regression(d.x_train,d.y_train,d.x_test,d.y_test); 
\end{lstlisting}

\noindent {\bf Step 3: Perform solver}

Now, you can perform optimization solvers, i.e., SGD and SVRG, calling solver functions, i.e., \verb+sgd()+ function and \verb+svrg()+ function, after setting some optimization options as the \verb+options+ struct.

\begin{lstlisting}[language=Matlab, label=demo_execute_solvers, caption=Code for optimization solver execution., firstnumber=7]
options.w_init = d.w_init;                      % set initial value  
options.step_init = 0.01;                       % set initial stepsize
opitons.verbose = 2;                            % set verbose mode
[w_sgd, info_sgd] = sgd(problem, options);      % perform SGD solver
[w_svrg, info_svrg] = svrg(problem, options);   % perform SVRG solver
\end{lstlisting}

They return the final solutions of \verb+w+ and the statistics information that include the histories of the epoch numbers, the cost function values, norms of gradient, the number of gradient evaluations and so on.

\noindent {\bf Step 4: Show result}

Finally, \verb+display_graph()+ provides output results of decreasing behavior of the cost values in terms of the number of gradient evaluations. Note that each algorithm needs different number of evaluations of samples in each epoch. Therefore, it is common to use this number to evaluate stochastic optimization algorithms instead of the number of iterations.

\begin{lstlisting}[language=Matlab, label=demo_show_results, caption=Code for showing results., firstnumber=13]
% display cost vs. number of gradient evaluations
display_graph('grad_calc_count','cost',{'SGD','SVRG'},...
                {w_sgd,w_svrg},{info_sgd,info_svrg});
\end{lstlisting}

\clearpage
\section{Problem definition}

This specifies the problem of interest with respect to w, noted as \verb+w+ in the library. It is noteworthy that the problem is defined as the class-based structure by exploiting MATLAB \verb+classdef+.  

The user does nothing other than calling a problem definition function. 

The build-in problems in the library include 
\begin{itemize}
	\item \L{2}-norm regularized multidimensional linear regression
	\item \L{2}-norm regularized linear support vector machines (SVM)
	\item \L{2}-norm regularized logistic regression (LR)
	\item \L{2}-norm regularized softmax classification (multinomial LR)
	\item \L{1}-norm regularized multidimensional linear regression	
	\item \L{1}-norm regularized logistic regression (LR)
\end{itemize}

Each problem definition contains the functions necessary for solvers;

\begin{itemize}
\item \verb+cost(w)+: calculate full cost function $f(w)$,
\item \verb+grad(w,indices)+ : 
calculate mini-batch stochastic derivative $\verb+v+=1/|\mathcal{S}| \nabla f_{i \in \mathcal{S}} (w)$ for the set of samples $\mathcal{S}$, which is noted as \verb+indices+.
\item \verb+hess(w,indices)+: 
calculate stochastic Hessian for \verb+indices+.
\item 
\verb+hess_vec(w,v,indices)+: 
calculate stochastic Hessian-vector product for \verb+v+ and \verb+indices+.
\end{itemize}

We illustrate the code of the definition of \L{2}-norm regularized linear regression function as an example. Listing \ref{list_lr_all} shows its entire code of \verb+linear_regression()+ function.

\begin{lstlisting}[language=Matlab, label=list_lr_all, caption=Demonstration code for linear regression problem.]
classdef linear_regression
    properties
        name;    
        dim;
        samples;
        lambda;
        hessain_w_independent;
        d;
        n_train;
        n_test;
        x_train;
        y_train;
        x_test;
        y_test;
        x_norm;
        x;         
    end
    
    methods
        % define constructor
        function obj=linear_regression(x_train, y_train, x_test, y_test, varargin) 
            obj.x_train = x_train;
            obj.y_train = y_train;
            obj.x_test = x_test;
            obj.y_test = y_test;            
            if nargin < 5; obj.lambda = 0.1; 
            else; obj.lambda = varargin{1}; end
            
            obj.d = size(x_train, 1);
            obj.n_train = length(y_train);
            obj.n_test = length(y_test);      
            obj.name = 'linear_regression';    
            obj.dim = obj.d;
            obj.samples = obj.n_train;
            obj.hessain_w_independent = true;
            obj.x_norm = sum(x_train.^2,1);
            obj.x = x_train;    
        end

        % define cost function
        function f = cost(obj, w)
            f = sum((w'*obj.x_train-obj.y_train).^2)/(2 * obj.n_train) ...
                + obj.lambda/2*w'*w;
        end

        % define stochastic gradient
        function g = grad(obj, w, indices)
            residual = w'*obj.x_train(:,indices) - obj.y_train(indices);
            g = obj.x_train(:,indices) * residual'/length(indices) ...
                + obj.lambda*w;
        end

        % define full gradient
        function g = full_grad(obj, w)
            g = obj.grad(w, 1:obj.n_train);
        end

        % define stochastic hessian
        function h = hess(obj, w, indices)
            h = 1/length(indices) * obj.x_train(:,indices) ...
                * (obj.x_train(:,indices)') + obj.lambda * eye(obj.d);
        end

        % define full hessian
        function h = full_hess(obj, w)
            h = hess(w, 1:obj.n_train);
        end

        % define stochastic hessian-vector
        function hv = hess_vec(obj, w, v, indices)
            hv = 1/length(indices) * obj.x_train(:,indices) ...
                * ((obj.x_train(:,indices)'*v)) + obj.lambda*v;
        end

        % define prediction function
        function p = prediction(obj, w)
            p = w' * obj.x_test;        
        end

        % define mse calculation function
        function e = mse(obj, y_pred)
            e = sum((y_pred-obj.y_test).^2)/(2 * obj.n_test);
        end
    end
end

\end{lstlisting}

We explain line-by-line of Listing \ref{list_lr_all} below; 

Listing \ref{list_lr_properties} defines the properties (i.e., class variables) \verb+properties+ of the class.

\begin{lstlisting}[language=Matlab, label=list_lr_properties, caption=Properties for linear regression problem class., firstnumber=2]
    properties
        name;    
        dim;
        samples;
        lambda;
        hessain_w_independent;
        d;
        n_train;
        n_test;
        x_train;
        y_train;
        x_test;
        y_test;
        x_norm;
        x;         
    end
\end{lstlisting}

Listing \ref{list_lr_const} defines the constructor \verb+linear_regression()+ of the class.

\begin{lstlisting}[language=Matlab, label=list_lr_const, caption=Constructor for linear regression problem class., firstnumber=20]
        % define constructor
        function obj=linear_regression(x_train, y_train, x_test, y_test, varargin) 
            obj.x_train = x_train;
            obj.y_train = y_train;
            obj.x_test = x_test;
            obj.y_test = y_test;            
            if nargin < 5; obj.lambda = 0.1; 
            else; obj.lambda = varargin{1}; end
            
            obj.d = size(x_train, 1);
            obj.n_train = length(y_train);
            obj.n_test = length(y_test);      
            obj.name = 'linear_regression';    
            obj.dim = obj.d;
            obj.samples = obj.n_train;
            obj.hessain_w_independent = true;
            obj.x_norm = sum(x_train.^2,1);
            obj.x = x_train;    
        end
\end{lstlisting}

Listing \ref{list_lr_cost} defines the cost calculation function \verb+cost()+ with respect to \verb+w+.

\begin{lstlisting}[language=Matlab, label=list_lr_cost, caption=Cost function code for linear regression problem class., firstnumber=40]
        % define cost function
        function f = cost(obj, w)
            f = sum((w'*obj.x_train-obj.y_train).^2)/(2 * obj.n_train) ...
                + obj.lambda/2*w'*w;
        end
\end{lstlisting}

Listing \ref{list_lr_grad} defines the stochastic gradient calculation function \verb+grad()+ with respect to \verb+w+ for  \verb+indices+. \verb+full_grad()+ calculates the full gradient estimation. 

\begin{lstlisting}[language=Matlab, label=list_lr_grad, caption=Gradient calculation function code for linear regression problem class., firstnumber=53]
        % define cost function
        function f = cost(obj, w)
            f = sum((w'*obj.x_train-obj.y_train).^2)/(2 * obj.n_train) ...
                + obj.lambda/2*w'*w;
        end

        % define stochastic gradient
        function g = grad(obj, w, indices)
            residual = w'*obj.x_train(:,indices) - obj.y_train(indices);
            g = obj.x_train(:,indices) * residual'/length(indices) ...
                + obj.lambda*w;
        end
\end{lstlisting}

Listing \ref{list_lr_hess} defines the stochastic Hessian calculation function \verb+hess()+ with respect to \verb+w+ for  \verb+indices+. \verb+full_hess()+ calculates the full Hessian estimation. 

\begin{lstlisting}[language=Matlab, label=list_lr_hess, caption=Hessian calculation function code for linear regression problem class., firstnumber=64]
        % define stochastic hessian
        function h = hess(obj, w, indices)
            h = 1/length(indices) * obj.x_train(:,indices) ...
                * (obj.x_train(:,indices)') + obj.lambda * eye(obj.d);
        end

        % define full hessian
        function h = full_hess(obj, w)
            h = hess(w, 1:obj.n_train);
        end

\end{lstlisting}

Listing \ref{list_lr_hess_vec} defines the stochastic Hessian-vector product calculation function \verb+hess_vec()+ with respect to \verb+w+ and a vector \verb+v+ for \verb+indices+.

\begin{lstlisting}[language=Matlab, label=list_lr_hess_vec, caption=Hessian-vector product calculation function code for linear regression problem class., firstnumber=70]
        % define stochastic hessian-vector
        function hv = hess_vec(obj, w, v, indices)
            hv = 1/length(indices) * obj.x_train(:,indices) ...
                * ((obj.x_train(:,indices)'*v)) + obj.lambda*v;
        end

\end{lstlisting}

The problem descriptor also provides some specific functions that are necessary for a particular problem. For example, the linear regression problem predicts the class for test data based on the model parameter that is trained by a stochastic optimization algorithm. Then, the final classification are calculated. Therefore, the regression problem equips the prediction and the mean squares error (MSE) calculation functions. 
\begin{itemize}
\item \verb+prediction(w)+
\item \verb+mse(y_pred)+
\end{itemize}
Listing \ref{list_lr_pred_acc} shows the prediction function \verb+prediction+ from the final solution \verb+w+, and the MSE calculation function \verb+mse+.

\begin{lstlisting}[language=Matlab, label=list_lr_pred_acc, caption=Prediction and MSE calculation code for linear regression problem class., firstnumber=75]
        % define prediction function
        function p = prediction(obj, w)
            p = w' * obj.x_test;        
        end

        % define mse calculation function
        function e = mse(obj, y_pred)
            e = sum((y_pred-obj.y_test).^2)/(2 * obj.n_test);
        end
\end{lstlisting}

Meanwhile, in case of  the logistic regression problem,  the final classification are calculated. Therefore, the regression problem equips the prediction and the prediction accuracy calculation functions. 
\begin{itemize}
\item \verb+prediction(w)+
\item \verb+accuracy(y_pred)+
\end{itemize}
Listing \ref{list_logreg_pred_acc} illustrates the binary-class prediction function \verb+prediction+ with the final solution \verb+w+, and prediction accuracy calculation function \verb+accuracy+ with the predicted classes.

\begin{lstlisting}[language=Matlab, label=list_logreg_pred_acc, caption=Prediction and prediction accuracy calculation code for logistric regression problem.]
    % define prediction function    
    Problem.prediction = @prediction;
    function p = prediction(w)
        p = sigmoid(w' * x_test);
        class1_idx = p>0.5;
        class2_idx = p<=0.5;         
        p(class1_idx) = 1;
        p(class2_idx) = -1;         
    end

    % define prediction accuracy calculation function       
    Problem.accuracy = @accuracy;
    function a = accuracy(y_pred)
        a = sum(y_pred == y_test)/n_test; 
    end
\end{lstlisting}

The last example is the softmax regression problem case. This case also contains the prediction function \verb+prediction+ and the prediction accuracy calculation function \verb+accuracy+  as shown in Listing \ref{list_softmax_pred_acc}.

\begin{lstlisting}[language=Matlab, label=list_softmax_pred_acc, caption=Prediction and prediction accuracy calculation code for softmax regression problem.]
    % define prediction function
    Problem.prediction = @prediction;
    function max_class = prediction(w)
        w_mat = reshape(w, [d n_classes]);        
        p = w_mat' * x_test;
        [~, max_class] = max(p, [], 1);
    end

    % define prediction accuracy calculation function    
    Problem.accuracy = @accuracy;
    function a = accuracy(class_pred)
        [~, class_test] = max(y_test, [], 1);
        a = sum(class_pred == class_test) / n_test; 
    end
\end{lstlisting}

\clearpage
\section{Solver definition}

This implements the main routine of the stochastic optimization solver. The optimization solver solves an optimization problem by calling the corresponding functions via the problem definition, such as \verb+cost(w)+, \verb+grad(w,indices)+, and possibly \verb+hess(w,indices)+. The final optimal solution \verb+w+ and the statistic information are returned. The latter is stored as the \verb+info+ struct. See also Appendix \ref{Sec:CollectStatisticsInformation}.

This section illustrates the definition of the SGD function \verb+sgd()+ as an example. The entire code of \verb+sgd()+ is shown in Listing \ref{sgd_code}.
\begin{lstlisting}[language=Matlab, label=sgd_code, caption=Demonstration code for SGD solver.]
function [w,infos]=sgd(problem, in_options)
    n=problem.samples();                                % number of samples

    % set options 
    local_options=[];
    options=mergeOptions(get_default_options(problem.dim()),local_options);   
    options=mergeOptions(options,in_options);  
    
    % initialize
    iter=0; epoch=0; grad_calc_count=0;
    w=options.w_init;
    num_of_bachces=floor(n/options.batch_size); 

    % store first statistics infos
    infos=store_infos(problem,w,options,[],epoch,grad_calc_count,0);
    
    % outer loop
    s_time = tic();
    while(optgap>options.tol_optgap) && (epoch<options.max_epoch)
        perm_idx = randperm(n);                         % permute samples

        % inner loop
        for j = 1 : num_of_bachces
            step=options.stepsizefun(iter,options);     % step-size
            
            s_index=(j-1)*options.batch_size+1;
            indice_j=perm_idx(s_index:s_index+options.batch_size-1);
            grad=problem.grad(w, indice_j);             % calculate gradient
            
            w=w-step*grad;                              % update w
            
            if isfield(problem,'prox')                  % proximal operator
                w=problem.prox(w, step);
            end  
            
            iter=iter+1;
        end
        
        grad_calc_count=grad_calc_count+num_of_bachces*options.batch_size;        
        epoch=epoch+1;

        % store statistics infos
        infos=store_infos(problem,w,options,infos,epoch,...
                                grad_calc_count,toc(s_time));        
    end
end
\end{lstlisting}

We explain line-by-line of Listing \ref{sgd_code} below.
 
We first set option parameters as a \verb+options+ struct. The solver merges the local default options \verb+local_options+, the default options \verb+get_default_options()+ that are commonly used in all solvers, and the input options \verb+in_options+. This is shown in Listing \ref{sgd_code_options}.
\begin{lstlisting}[language=Matlab, label=sgd_code_options, caption=Optional parameter configuration code for SGD solver., firstnumber=4]
    % set options 
    local_options=[];
    options=mergeOptions(get_default_options(problem.dim()),local_options);   
    options=mergeOptions(options,in_options);  
\end{lstlisting}

Next, the initial statistics data are collected by \verb+store_infos+ function as shown in Listing \ref{sgd_code_info} before entering the main optimization routine. The statistics data are also collected at the end of every epoch, i.e., outer iteration, as Listing \ref{sgd_code_info_end}. See also Appendix \ref{Sec:CollectStatisticsInformation}.

\begin{lstlisting}[language=Matlab, label=sgd_code_info, caption=Initial statistics data collection code for SGD solver., firstnumber=113]
    % store first statistics infos
    infos=store_infos(problem,w,options,[],epoch,grad_calc_count,0);      
\end{lstlisting}

\begin{lstlisting}[language=Matlab, label=sgd_code_info_end, caption=Statistics data collection code for SGD solver., firstnumber=141]
        % store statistics infos
        infos=store_infos(problem,w,options,infos,epoch,...
                                grad_calc_count,toc(s_time));        
\end{lstlisting}

The code in Listing \ref{sgd_code_set_step_alg} shows the calculation of the stepsize. If the user does not specify his/her user-defined stepsize algorithm, the solver calls a default stepsize algorithm function with  \verb+options+ parameter. This parameter contains the stepsize algorithm type, the initial stepsize and so on. Otherwise, the user-defined stepsize algorithm is called.  See also Appendix \ref{Sec:UserDefStepsizeAlgorithm}.
\begin{lstlisting}[language=Matlab, label=sgd_code_set_step_alg, caption=Stepsize calculation code for SGD solver., firstnumber=24]
            step=options.stepsizefun(iter,options);     % step-size
\end{lstlisting}

Listing \ref{sgd_code_grad} illustrates the code for the stochastic gradient calculation, where \verb+indice_j+ is calculated from the number of the inner iteration, \verb+j+, and the batch size \verb+options.batch_size+.

\begin{lstlisting}[language=Matlab, label=sgd_code_grad, caption=Stochastic gradient calculation code for SGD solver., firstnumber=26]
            s_index=(j-1)*options.batch_size+1;
            indice_j=perm_idx(s_index:s_index+options.batch_size-1);
            grad=problem.grad(w, indice_j);             % calculate gradient
\end{lstlisting}

Finally, the model parameter \verb+w+ is updated using the calculated stepsize \verb+step+ and the stochastic gradient \verb+grad+ as Listing \ref{sgd_code_w_update}.

\begin{lstlisting}[language=Matlab, label=sgd_code_w_update, caption=Mode parameter update code for SGD solver., firstnumber=30]
            w=w-step*grad;                              % update w
\end{lstlisting}

\clearpage
\section{Default stepsize algorithms}
\label{Sec:DefaultStepsizeAlgorithms}

SGDLibrary supports four stepsize algorithms. This can be switched by the setting option struct such as \verb+options.step_alg='decay-2'+. After $\eta_0(=\verb+options.step_init+)$ and $\lambda(=\verb+options.lambda+)$ are properly configured, we can use one of the following algorithms according to the total inner iteration number $k$ as;
\begin{itemize}
	\item \verb+fix+: This case uses below;
	\begin{eqnarray*}
		\eta &=& \eta_0.
	\end{eqnarray*}	
	\item \verb+decay+	: This case uses below;
	\begin{eqnarray*}
		\eta &=& \frac{\eta_0}{1 + \eta_0 \lambda k}.
	\end{eqnarray*}
	\item \verb+decay-2+: This case uses below;
	\begin{eqnarray*}
		\eta &=& \frac{\eta_0}{1 + k}.
	\end{eqnarray*}		
	\item \verb+decay-3+: This case uses below;
	\begin{eqnarray*}
		\eta &=& \frac{\eta_0}{\lambda + k}.
	\end{eqnarray*}			
\end{itemize}

SGDLibrary also accommodates a user-defined stepsize algorithm. See the next section.

\clearpage
\section{User-defined stepsize algorithm}
\label{Sec:UserDefStepsizeAlgorithm}

SGDLibrary allows the user to define a new user-defined stepsize algorithm. This is done by setting as \verb+options.stepsizefun = @my_stepsize_alg+, which is delivered to solvers via their input arguments. The illustrative example code is shown in Listing \ref{user_define_step_alg}.

\begin{lstlisting}[language=Matlab, label=user_define_step_alg, caption=Demonstration code for user-define stepsize algorithm.]
function stepsize_alg_demo()
    % generate synthetic 1000 samples of dimension 10     
    d = logistic_regression_data_generator(1000,10);
        
    % define logistic regression problem
    problem = logistic_regression(d.x_train, d.y_train, d.x_test, d.y_test); 
    
    options.w_init = d.w_init;          % set initial value    
    options.step_init = 0.01;           % set initial stepsize  
    options.step_alg = 'decay';         % set decay stepsize algorithm
    [w_sgd,info_sgd] = sgd(problem, options);      % perform SGD solver    
    options.stepsizefun = @my_stepalg;  % set my_stepalg stepsize algorithm
    [w_sgd_my,info_sgd_my] = sgd(problem, options);% perform SGD solver 
    
    % display cost/optimality gap vs number of gradient evaluations
    display_graph('grad_calc_count','cost',{'SGD(decay)','SGD(My algorithm)'},{w_sgd, w_sgd_my},{info_sgd,info_sgd_my});
end

% define user-defined stepsize algorithm
function step = my_stepalg(iter,options)
    step = options.step_init/(10+iter*0.5);
end
\end{lstlisting}

Listing \ref{user_define_step_alg_def} defines an example of the user-defined stepsize algorithm named as \verb+my_stepalg+.
\begin{lstlisting}[language=Matlab, label=user_define_step_alg_def, caption=Definition code of user-define stepsize algorithm., firstnumber=19]
% define user-defined stepsize algorithm
function step = my_stepalg(iter,options)
    step = options.step_init/(10+iter*0.5);
end

\end{lstlisting}

Then, the new algorithm function \verb+my_stepalg+ is set to the algorithm via \verb+options+ value for each solver as shown in Listing \ref{user_define_step_alg_set}.
\begin{lstlisting}[language=Matlab, label=user_define_step_alg_set, caption=Setting code of user-define stepsize algorithm., firstnumber=12]
    options.stepsizefun = @my_stepalg;  % set my_stepalg stepsize algorithm


\end{lstlisting}

\clearpage
\section{Collect statistics information}
\label{Sec:CollectStatisticsInformation}

The solver automatically collects statistics information every epoch, i.e., outer iteration, via  \verb+store_infos()+ function. The collected data are stored in \verb+infos+ struct, and it is returned to the caller function to visualize the results. The information contain below.
\begin{itemize}
    \item \verb+iter+: number of iterations
    \item \verb+time+: elapsed time
    \item \verb+grad_calc_count+: count of gradient calculations
    \item \verb+optgap+: optimality gap
    \item \verb+cost+: cost function value
    \item \verb+gnorm+: norm of full gradient
\end{itemize}

Additionally, when a regularizer exists, its value is collected. This provides informative information, for example, the sparsity when $R(w)=\| w\|_1$. Furthermore, the user can collect the history of solution \verb+w+ in every epoch. 

The code of \verb+store_infos()+ is illustrated in Listing \ref{lstore_infos}.
\begin{lstlisting}[language=Matlab, label=lstore_infos, caption=Statistic information collection code.]
function [infos,f_val,optgap] = store_infos(problem,w,options,infos,...
                                    epoch,grad_calc_count,elapsed_time)
    % calculation of cost function value
    f_val = problem.cost(w);
    % calculation of norm of full gradient
    gnorm = norm(problem.full_grad(w));  
    % calculation of optimality gap
    optgap = f_val - options.f_opt;     
        
    % number of iterations
    infos.iter = [infos.iter epoch];
    % elapsed time
    infos.time = [infos.time elapsed_time];
    % count of gradient calculations
    infos.grad_calc_count = [infos.grad_calc_count grad_calc_count];
    % optimality gap
    infos.optgap = [infos.optgap optgap];
    % cost function value
    infos.cost = [infos.cost f_val];
    % norm of full gradient
    infos.gnorm = [infos.gnorm gnorm]; 
    % value of regularizer
    if isfield(problem, 'reg')
        reg = problem.reg(w);
        infos.reg = [infos.reg reg];
    end   
    % solution w
    if options.store_w
        infos.w = [infos.w w];         
    end  
end

\end{lstlisting}

\clearpage
\section{Visualizations}

SGDLibrary provides various visualization tools, which include
\begin{itemize}
	\item \verb+display_graph()+: display various graphs such as cost function values vs. the number of gradient evaluations, and the optimality gap vs. the number of gradient evaluations.
	\item \verb+display_regression_result()+: show regression results for regression problems. 
	\item \verb+display_classification_result()+: show classification results specialized for classification problems.
	\item \verb+draw_convergence_animation()+: draw convergence behavior animation of solutions. 
\end{itemize}

This section provides illustrative explanations in case of \L{2}-norm regularized logistic regression problem. The entire code is shown in Listing \ref{visualization}.
\begin{lstlisting}[language=Matlab, label=visualization, caption=Demonstration code for result visualization.]
% generate synthetic 300 samples of dimension 2 for logistic regression
d=logistic_regression_data_generator(300,2);

% define problem definitions
problem=logistic_regression(d.x_train,d.y_train,d.x_test,d.y_test); 

w_opt=problem.calc_solution(problem,1000);          % calculate optimal w
options.f_opt=problem.cost(w_opt);                  % calculate optimal f
options.store_w=true;                  % store w for convergence animation

options.w_init=d.w_init;                            % set initial value  
options.step_init=0.01;                             % set initial stepsize
[w_sgd,info_sgd]= gd(problem,options);              % perform SGD solver
[w_svrg,info_svrg]=svrg(problem,options);           % perform SVRG solver

% display cost/optimality gap vs number of gradient evaluations
display_graph('grad_calc_count','cost',{'SGD','SVRG'},...
                {w_sgd,w_svrg},{info_sgd,info_svrg});
display_graph('grad_calc_count','optimality_gap',{'SGD','SVRG'},...
                {w_sgd,w_svrg},{info_sgd,info_svrg});    

% display classification results 
y_pred_sgd=problem.prediction(w_sgd);       % predict for SGD
accuracy_sgd=problem.accuracy(y_pred_sgd);  % calculate accuracy for SGD
fprintf('Classificaiton accuracy:%s:%.4f\n','SGD',accuracy_sgd);
y_pred_sgd(y_pred_sgd==-1)=2;               % convert from {1,-1} to {1,2}
y_pred_sgd(y_pred_sgd==1)=1; 
y_pred_svrg = problem.prediction(w_svrg);   % predict for SVRG
accuracy_svrg=problem.accuracy(y_pred_svrg);% calculate accuracy for SVRG
fprintf('Classificaiton accuracy:%s:%.4f\n','SVRG',accuracy_svrg);
y_pred_svrg(y_pred_svrg==-1) = 2;           % convert from {1,-1} to {1,2}
y_pred_svrg(y_pred_svrg==1) = 1;
d.y_train(d.y_train==-1)=2; d.y_train(d.y_train==1)=1;
d.y_test(d.y_test==-1)=2;   d.y_test(d.y_test==1)=1;  
display_classification_result(problem, {'SGD','SVRG'},{w_sgd,w_svrg}, ...
                {y_pred_sgd,y_pred_svrg},{accuracy_sgd,accuracy_svrg}, ...
                d.x_train,d.y_train,d.x_test,d.y_test);    

% display convergence animation
draw_convergence_animation(problem,{'SGD','SVRG'},...
                {info_sgd.w,info_svrg.w},100,0.1);        
\end{lstlisting}

First, for the calculation of {\it optimality gap}, the user needs the optimal solution \verb+w_opt+ beforehand. This is obtained by calling \verb+problem.calc_solution()+ function of the problem definition function. This case uses the L-BFGS solver inside it to obtain the optimal solution under maximum iteration $1000$ with a very precise tolerant stopping condition. Then, the optimal cost function value \verb+f_opt+ is calculated from \verb+w_opt+. Listing \ref{visualization_w_opt} shows the code. 
\begin{lstlisting}[language=Matlab, label=visualization_w_opt, caption=Optimal solution and cost function value calculation code., firstnumber=7]
w_opt=problem.calc_solution(problem,1000);          % calculate optimal w
options.f_opt=problem.cost(w_opt);                  % calculate optimal f       
\end{lstlisting}

Then, you obtain the result of the optimality gap by \verb+display_graph()+ as Listing \ref{visualization_optgap}. The first argument and the second argument correspond to the values of $x$-axis and $y$-axis, respectively, in the graph. We can change these values according to the statistics data explained in Appendix \ref{Sec:CollectStatisticsInformation}. The third argument represents the list of the algorithm names that are shown in the legend of the graph. The forth parameter indicates the list of the \verb+info+ structs to be shown. 
\begin{lstlisting}[language=Matlab, label=visualization_optgap, caption=Display graph code., firstnumber=16]
% display cost/optimality gap vs number of gradient evaluations
display_graph('grad_calc_count','cost',{'SGD','SVRG'},...
                {w_sgd,w_svrg},{info_sgd,info_svrg});
display_graph('grad_calc_count','optimality_gap',{'SGD','SVRG'},...
                {w_sgd,w_svrg},{info_sgd,info_svrg});    
\end{lstlisting}

Additionally, in case of \L{2}-norm regularized logistic regression problems, the results of classification accuracy are calculated using the corresponding prediction function \verb+probrem.prediction()+ and the accuracy calculating function \verb+probrem.accuracy()+. Then, the classification accuracies are illustrated by \verb+display_classification_result()+ function. The code is shown as Listing \ref{visualization_pred}.
\begin{lstlisting}[language=Matlab, label=visualization_pred, caption=Display classification results code., firstnumber=22]
% display classification results 
y_pred_sgd=problem.prediction(w_sgd);       % predict for SGD
accuracy_sgd=problem.accuracy(y_pred_sgd);  % calculate accuracy for SGD
fprintf('Classificaiton accuracy:%s:%.4f\n','SGD',accuracy_sgd);
y_pred_sgd(y_pred_sgd==-1)=2;               % convert from {1,-1} to {1,2}
y_pred_sgd(y_pred_sgd==1)=1; 
y_pred_svrg = problem.prediction(w_svrg);   % predict for SVRG
accuracy_svrg=problem.accuracy(y_pred_svrg);% calculate accuracy for SVRG
fprintf('Classificaiton accuracy:%s:%.4f\n','SVRG',accuracy_svrg);
y_pred_svrg(y_pred_svrg==-1) = 2;           % convert from {1,-1} to {1,2}
y_pred_svrg(y_pred_svrg==1) = 1;
d.y_train(d.y_train==-1)=2; d.y_train(d.y_train==1)=1;
d.y_test(d.y_test==-1)=2;   d.y_test(d.y_test==1)=1;  
display_classification_result(problem, {'SGD','SVRG'},{w_sgd,w_svrg}, ...
                {y_pred_sgd,y_pred_svrg},{accuracy_sgd,accuracy_svrg}, ...
                d.x_train,d.y_train,d.x_test,d.y_test);  
\end{lstlisting}

\begin{figure}
\begin{center}
\includegraphics[width=15cm]{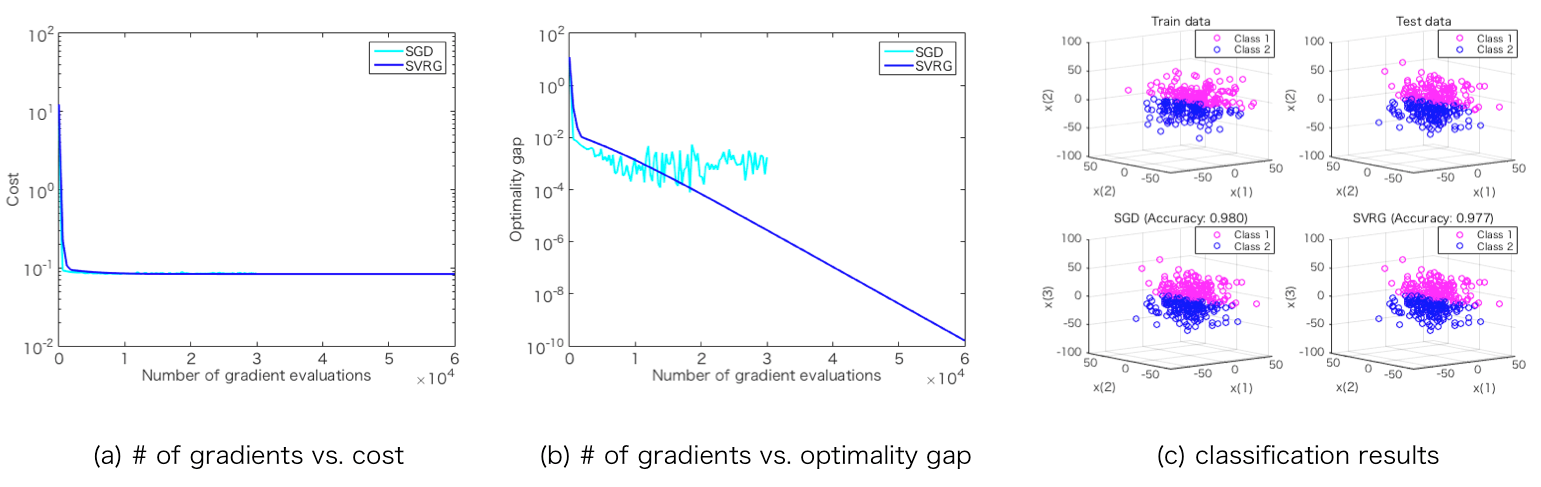}
\caption{Logistic regression problem.} 
 \label{sample-figure}
\end{center}
\end{figure}

Finally, you can also show a demonstration of {\it convergence animation}. You need specify additional options before executing solvers as Listing \ref{visualization_store_w_flag}.
\begin{lstlisting}[language=Matlab, label=visualization_store_w_flag, caption=Store history of solutions., firstnumber=9]
options.store_w=true;                  % store w for convergence animation
\end{lstlisting}

Now, the animation of the convergence behavior is shown. The code is shown in Listing \ref{visualization_conv_anime}. It should be noted that \verb+draw_convergence_animation()+ is executable when only the dimension $d$ of the parameters is $2$. The last parameter for the function, i.e., $0.1$ in this example, indicates the speed of the animation.  
\begin{lstlisting}[language=Matlab, label=visualization_conv_anime, caption=Draw convergence animation., firstnumber=39]
% display convergence animation
draw_convergence_animation(problem,{'SGD','SVRG'},...
                {info_sgd.w,info_svrg.w},100,0.1);        
\end{lstlisting}

\begin{figure}[htbp]
\begin{center}
\includegraphics[width=15cm]{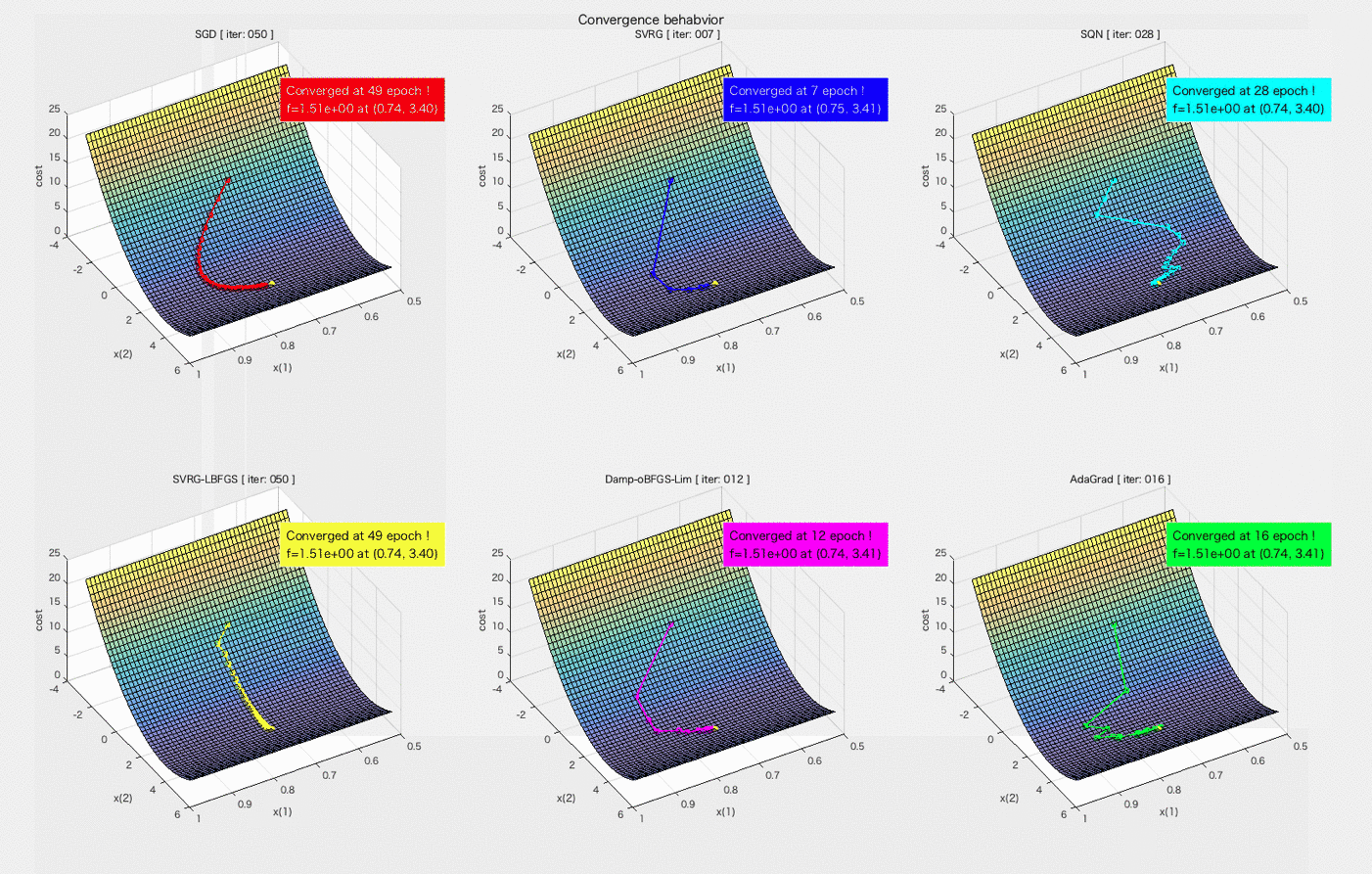}
\caption{convergence animation.} 
 \label{sample-figure}
\end{center}
\end{figure}

\clearpage
\section{More results}

We show more results for 
\L{2}-norm regularized linear regression problem, 
\L{2}-norm softmax classifier problem, and 
\L{2}-norm regularized linear SVM problem in Figures \ref{Results_linear_reg} to \ref{Results_SVM_reg}, respectively. 

\begin{figure}[htbp]
\begin{center}
\includegraphics[width=15cm]{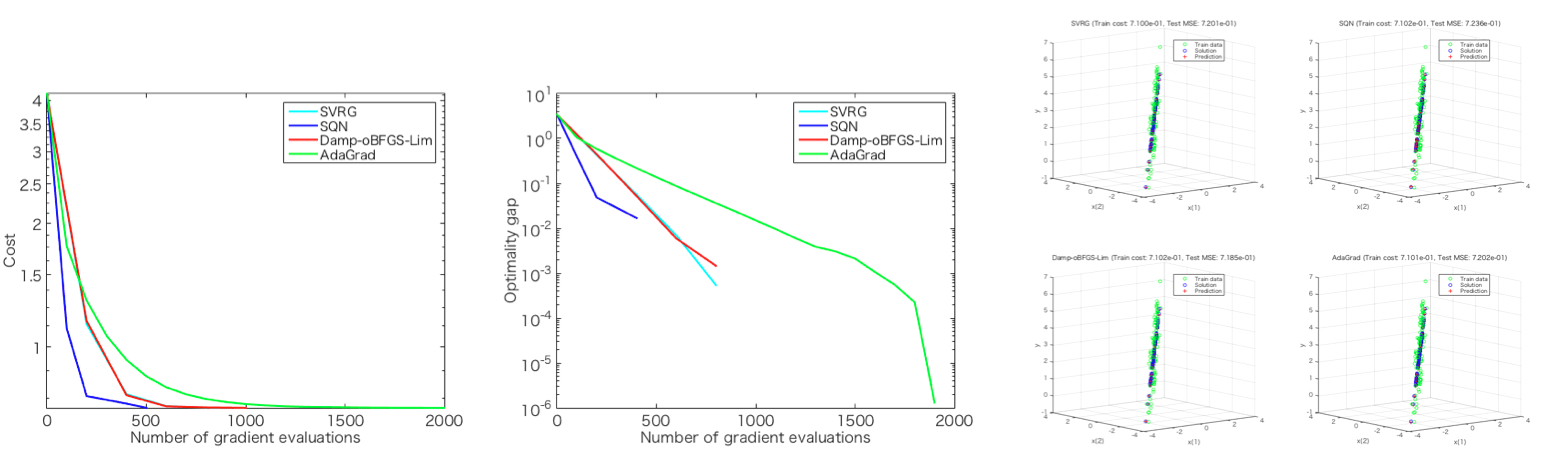}
\caption{Linear regression problem.} 
\label{Results_linear_reg}
\end{center}
\end{figure}

\begin{figure}[htbp]
\begin{center}
\includegraphics[width=15cm]{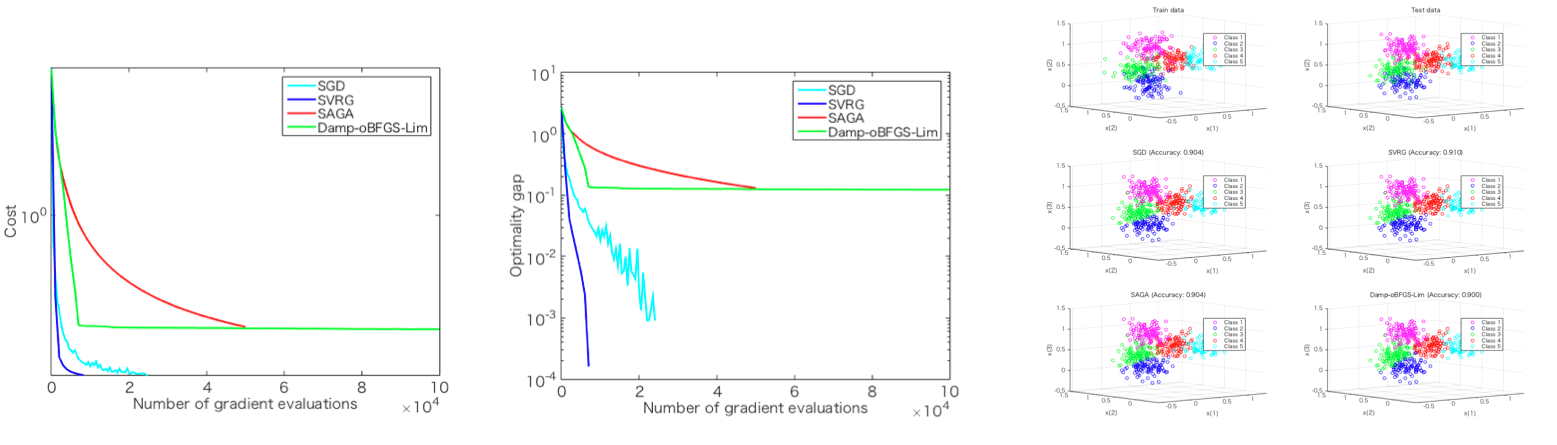}
\caption{Softmax classifier problem.} 
\label{Results_softmax_reg}
\end{center}
\end{figure}

\begin{figure}[htbp]
\begin{center}
\includegraphics[width=15cm]{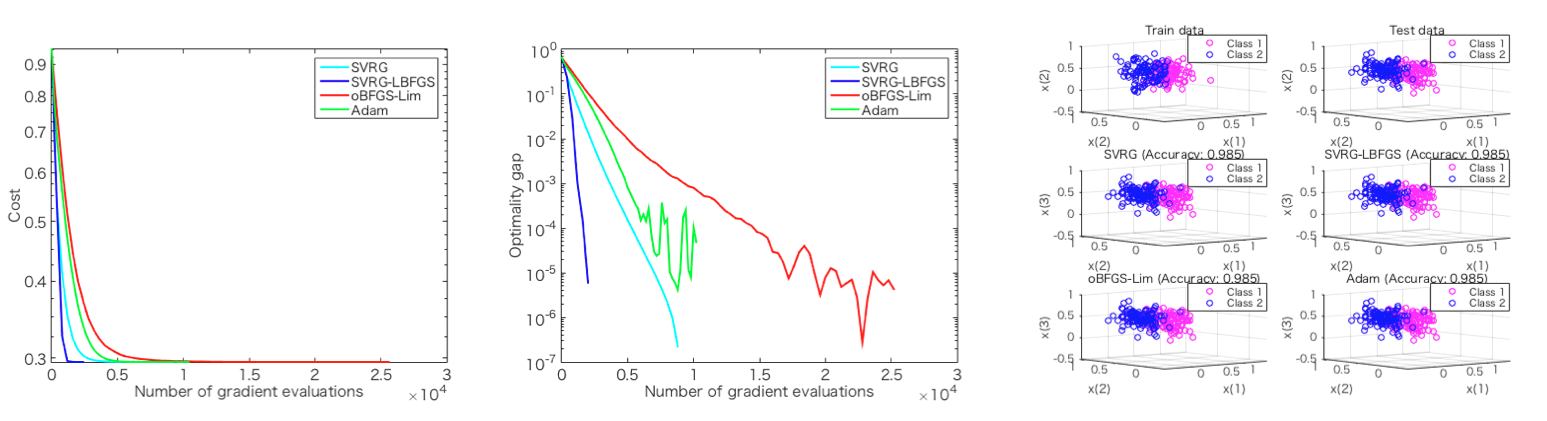}
\caption{Linear SVM problem.} 
\label{Results_SVM_reg}
\end{center}
\end{figure}

\clearpage
\bibliographystyle{unsrt}
\bibliography{/Users/kasai/Dropbox/DOC/Research/bibtex/stochastic_online_learning}

\end{document}